# Generating geographically and economically realistic large-scale synthetic contact networks: A general method using publicly available data

Alexander Y. Tulchinsky[1*], Fardad Haghpanah[1], Alisa Hamilton[1,2], Nodar Kipshidze[1], Eili Y. Klein[1,3,4], for the CDC MIND-Healthcare Program[^]

[1]One Health Trust, Washington, DC, United States of America
[2]Center for Systems Science and Engineering, Johns Hopkins University, Baltimore, MD, USA
[3]Department of Emergency Medicine, Johns Hopkins School of Medicine, Baltimore, MD, USA
[4]Department of Infectious Disease, Johns Hopkins School of Medicine, Baltimore, MD, USA
[^]Membership of the CDC MIND-Healthcare Program is provided at https://www.cdc.gov/hai/research/MIND-Healthcare.html

[*]Corresponding author
E-mail: tulchinsky@onehealthtrust.org

## Abstract

Synthetic contact networks are useful for modeling epidemic spread and social transmission, but data to infer realistic contact patterns that take account of assortative connections at the geographic and economic levels is limited. We developed a method to generate synthetic contact networks for any region of the United States based on publicly available data. First, we generate a synthetic population of individuals within households from US census data using combinatorial optimization. Then, individuals are assigned to workplaces and schools using commute data, employment statistics, and school enrollment data. The resulting population is then connected into a realistic contact network using graph generation algorithms. We test the method on two census regions and show that the synthetic populations accurately reflect the source data. We further show that the contact networks have distinct properties compared to networks generated without a synthetic population, and that those differences affect the rate of disease transmission in an epidemiological simulation. We provide open-source software to generate a synthetic population and contact network for any area within the US.

## Introduction

Synthetic contact networks are a valuable tool for modeling epidemic spread[1], opinion change[2], economic transactions[3], and other forms of transmission in studies of social systems. A realistic contact network, informed by individual demographic characteristics and contact patterns, is especially desirable for addressing questions of practical interest in a real-world population. In epidemiological research, these include topics such as testing the effect of detailed interventions (e.g., closing schools, or restricting mobility) in which specific contact types are altered or examining disparities in health outcomes associated with race and income, or testing ideas relevant to epidemic forecasting. For example, networks favoring repeated contact among the same individuals were shown to produce slower disease spread compared to predictions based on full mixing[4,5]. Higher rates of incidental contact, "super-spreader" events, or periodic disruption/rewiring of the network (e.g., due to holiday travel) would be expected to increase mixing and produce place- and time- specific increases in infections.

Data to directly infer social contacts in real populations are limited or unavailable. To generate a realistic synthetic contact network, first a geographically realistic population needs to be generated. Generating a synthetic population with realistic placement and activities of individuals is challenging; detailed information is needed on population features such as household composition, school attendance, employment locations, and commute patterns. Census data provides information about geography and demographics at a suitably fine scale, but publicly available data are restricted to select summaries of marginal counts (e.g., how many households in a locale have five members and how many have school-age children) and are missing compositional details (e.g., the number of five-member households with school-age children). The marginal counts are generally not independent, and information about their joint distributions is not directly provided.

A variety of methods have been proposed for reconstructing individuals and households from aggregate census counts (reviewed in [6,7]). Commonly, these methods make use of microdata survey samples containing the quantities of interest in disaggregated form, which allows their approximate joint distributions to be recovered for each region covered by the microdata. In the United States, Public Use Microdata Samples (PUMS) are an anonymized subset of census survey responses for regions containing around 100,000 people. In this study, we adapted a relatively simple and well-performing method which uses combinatorial optimization (CO) to sample households from PUMS until a set is found that sufficiently matches the marginal counts from the fine-scale census data[8]. We then integrated those results with other data sources, including commute patterns, workplace characteristics, and school locations to generate location-bound activities outside of the household that are consistent with the attributes of the sampled individuals (e.g., [9,10]). By doing so, we produced a geographically detailed synthetic population reflecting the demographics, school attendance, and workplace composition of a chosen region, along with a corresponding network of regular household, school, and workplace contacts. We tested the method for two different regions of the US: the Maryland / Washington, D.C. / Northern Virginia region, and the five counties containing the greater Los Angeles area. We have made available a software package, GREASYPOP-CO, to automatically generate a synthetic population and associated contact network for any area of the US. The software and source code are available at https://github.com/CDDEP-DC/GREASYPOP-CO.

**Methods**

*Data sources*
From the US census, we obtained data at the census block group (CBG) level related to demography, household composition, employment, and group quarters (GQ) residency. We used American Community Survey (ACS) 5-year summary data from 2019 and decennial census data from 2010 (the most recent data having the same geographic boundaries as the 2019 ACS). The tables used are summarized in Table 1. In addition, we obtained the geographic location of each CBG from the 2019 TIGER/Line shapefiles provided by the US census.

We obtained the 2019 5-year PUMS from the US census at the household and individual levels for states relevant to the area being synthesized. For example, to synthesize the Maryland/DC region, we used samples from MD, DC, VA, and neighboring east coast states. For the Los Angeles area, we used samples from the entirety of California.

For work commute patterns, we used origin-destination (OD) and workplace area characteristics (WAC) data from Longitudinal Employer-Household Dynamics (lehd.ces.census.gov), version 7 (having the same geographic boundaries as the 2019 ACS), primary jobs file (JT01). We obtained main and auxiliary OD files for any state overlapping the synthesis area, and additional auxiliary files for states within likely commuting distance. To estimate employer sizes, we used the 2016 County Business Patterns (CBP) dataset (www.census.gov/programs-surveys/cbp.html).

From the National Center for Education Statistics (nces.ed.gov), we obtained a directory of US public schools, including student and staff counts, from the 2018-2019 Common Core of Data (CCD) dataset, and geographic coordinates from the 2018-2019 EDGE dataset. From the Missouri Census Data Center (mcdc.missouri.edu), we used the Geocorr 2018 dataset to obtain the urban/rural proportion for each CBG and PUMS sampling area.

**Table 1. Data tables obtained from American Community Survey and US decennial census.**

| Table | Description |
|---|---|
| B01001 | Sex by Age |
| B09018 | Relationship to Householder for Children Under 18 Years in Households |
| B09019 | Household Type (Including Living Alone) by Relationship |
| B09020 | Relationship by Household Type (Including Living Alone) for the Population 65 Years and Over |
| B09021 | Living Arrangements of Adults 18 Years and Over by Age |
| B11001H | Household Type, Householder is White Alone, Not Hispanic or Latino |
| B11001I | Household Type, Householder is Hispanic or Latino |
| B11004 | Family Type by Presence and Age of Related Children Under 18 Years |
| B11012 | Households by Type |
| B11016 | Household Type by Household Size |
| B19001 | Household Income in the Past 12 Months |
| B22010 | Receipt of Food Stamps/Snap in the Past 12 Months by Disability Status for Households |
| B23009 | Presence of Own Children Under 18 Years by Family Type by # of Workers in Family |
| B23025 | Employment Status for the Population 16 Years and Over |
| B25006 | Race of Householder |
| C24030 | Industry for Civilian Employed Population 16 Years and Over |
| P43 | Group Quarters Population by Sex by Age by Group Quarters Type |

*Preprocessing*

To avoid overly granular targets for optimization, we combined census data columns as follows. In ACS tables that count householders (B23009, B11004, and B11012), we combined male and female single householders. Additionally, we combined counts of married and unmarried/cohabiting households (B11012) and individuals (B09021). We combined household income categories (B19001) into eight groups: three for the lowest income quintile plus quintiles two through five, mapping population income quintiles as closely as possible to the category boundaries in the census table. The extra granularity in the lowest quintile was intended to distinguish households with low-income full-time wage earners from those without full-time workers. We combined the industry categories reported in table C24030 into 15 distinct categories (Table 2).

From the census columns, we selected targets whose counts would be matched with PUMS household samples using CO search. Targets were selected for their relevance to disease transmission, and their utility in generating household, school, and workplace contact networks. We retained a total of 85 target counts, covering the individual and household attributes summarized in Table 3. We joined household and person level PUMS data, and calculated household-level sums corresponding to each targeted census column, so that sets of sample households could easily be compared to census totals.

For each CBG, we calculated the total number of people in group quarters by age category (under 18, 18 to 64, and 65 and over) based on the total number of adults from ACS table B01001, adults in households from table B09021, total GQ population from table B09019, and people aged 65 and over in GQs from table B09020. For each age category, we then derived estimated counts of people in institutional GQs (e.g., prisons and nursing homes), civilian non-institutional GQs (e.g., college dorms), and military quarters using proportions based on the counts in decennial census table P43.

Because the CBG-level employment data (census tables C24030 and B23025) do not differentiate individuals by residence type (households vs. group quarters), we used iterative proportional fitting (IPF; [11]) to estimate counts of workers by industry crossed with residence type. For each CBG, the column sums were constrained to equal the census employment counts by industry category, and row sums were constrained to equal the number of individuals by residence type (households, non-institutional civilian GQ, or military GQ). Initial cell values were set as follows. For households, census employment counts were multiplied by the proportion of people in households for that CBG.

**Table 2. Industry categories derived from census for synthetic population construction.**

| Category | NAICS | Census columns |
| --- | --- | --- |
| AGR_EXT | 11 21 | Agriculture forestry fishing and hunting and mining |
| CON | 23 | Construction |
| MFG | 31 32 33 | Manufacturing |
| WHL | 42 | Wholesale trade |
| RET | 44 45 | Retail trade |
| TRN_UTL | 48 49 22 | Transportation and warehousing and utilities |
| INF | 51 | Information |
| FIN | 52 53 | Finance and insurance and real estate and rental and leasing |
| PRF | 54 55 56 | Professional scientific and management and administrative and waste management services |
| EDU | 61 | Educational services and health care and social assistance:Educational services |
| MED | 62 | Educational services and health care and social assistance:Health care and social assistance |
| ENT_art | 71 | Arts entertainment and recreation and accommodation and food services:Arts entertainment and recreation |
| ENT_food | 72 | Arts entertainment and recreation and accommodation and food services:Accommodation and food services |
| SRV | 81 | Other services except public administration |
| ADM_MIL | 92 | Public Administration, Armed Forces |

For civilian GQs, we estimated the proportion of civilian GQ workers in each industry category from state-level PUMS data, assumed that only non-institutionalized GQ residents aged 18 to 64 are employed, and multiplied the number of residents by the estimated industry proportions. For military GQs, all residents were initially assumed to work in the armed forces. IPF converged to a solution for each CBG in our test populations.

For generating the synthetic population, we kept only those CBGs having either 20 or more households or 20 or more GQ residents. In addition, we excluded a small number of CBGs that lacked a corresponding microdata sampling area (PUMA) in the 2010 census tract to 2010 PUMA relationship file provided by the US census.

Using the OD employment statistics from LEHD, which provide a count of work commuters between every pair of census blocks, we generated a proportional origin-destination commute matrix at the CBG level. That matrix was then used as the basis for creating an industry-by-destination commute matrix for each origin CBG, which was done using IPF as follows. For each CBG, row sums were constrained to equal the number of workers in each industry category based on census counts. Column sums were constrained to equal the total number of workers going to each destination, obtained by multiplying the total number of workers in that CBG by the corresponding row from the origin-destination matrix. Initial cell values were set using the proportion of workers in each industry by destination, obtained from WAC data, multiplied by the number of workers going to each destination. Thus, information on the industry composition of each work destination would be incorporated, while IPF would force the total number of workers to agree with the industry-specific census counts and the origin-destination data. IPF converged to a solution for each CBG in our test populations.

Work destinations outside the synthesis area were treated as a single location, as workplace networks will not be generated for those. Home origins outside the synthesis area were likewise treated as a single location, as household networks will not be generated for those commuters. For each county in the synthesis area, we assumed a log-normal distribution of employer sizes and used the counts of employers by size category in the CBP to estimate $\mu$ and $\sigma$ parameters for each distribution. The commute matrices and employer sizes were used to generate workplace contact networks after using CO to generate workers living in households in each CBG (see below for details).

**Table 3. Individual and household attributes selected for optimization search.**

| Table | Targeted counts | # of targets |
|---|---|---|
| B11016 | # of family and non-family households by size | 13 |
| B23009 | # of workers per family, by married/unmarried and presence of children under 18 (workers defined as people having worked in the past 12 months) | 16 |
| B11004 | # of families by presence and age group of related children, and by married/unmarried status | 8 |
| B09018 | # of children and grandchildren in households | 2 |
| B11012 | # two-partner households with and without children; single householders alone, with children under 18, with other relatives, and with non-relatives | 6 |
| B09021 | # of adults by age group and living arrangements (alone/with partner/with parent/with other relatives/with non-relatives) | 14 |
| B19001 | # of households be income category | 7 |
| B25006 | # of householders who are Black or African American alone | 1 |
| B11001H/I | # of White non-Hispanic householders, and # of Hispanic householders | 2 |
| C24030/B23025 | # of individuals employed by industry, including armed forces | 15 |
| B22010 | # of households receiving Food Stamps/SNAP | 1 |

We filtered the NCES public school data to include only active schools, in the "regular schools" category, with reported grade levels. For schools missing counts of students or teachers, we filled the missing values with the mean for that school type in the synthesis area. For assigning students to schools, we pre-computed the five closest public schools for each CBG at each grade level, using the geographic centroid of each CBG calculated in 18N projection coordinates (and converting the latitude-longitude coordinates of each school to the same projection). The school data were used to generate within-school contact networks after using CO to generate PK-12 students in each CBG.

*Synthetic population and network generation*

We adapted the combinatorial optimization method described in [8] to develop a population of geographically situated households from fine-scale census data. We selected CO for its ease of use and its empirically demonstrated accuracy in recreating properties of a known population[12,13]. To populate each CBG with households, we performed a CO search using simulated annealing to find the microdata (PUMS) samples best matching the census target columns. We started with a random selection of sample households, and randomly replaced one at each time step, accepting or rejecting the replacement with a probability based on the Freeman-Tukey ($FT^2$) goodness-of-fit statistic (following [13]) and the annealing temperature as follows. The acceptance probability is given by

$$P(accept) = \begin{cases} \exp(-\Delta E/T), & \Delta E \geq 0 \\ 1, & \Delta E < 0, \end{cases} \quad (1)$$

where $T$ is annealing temperature and $E$ is the mismatch cost, calculated as ¼ the $FT^2$ statistic:

$$E = \frac{FT^2}{4} = \sum_i \left(\sqrt{o_i} - \sqrt{e_i}\right)^2, \quad (2)$$

where $o_i$ and $e_i$ are the values of the $i$-th target count from the synthetic population and the census. We used ½ the mismatch cost of the initial selection as the starting temperature and decreased the temperature at each time step by a constant multiplier (0.99 for the first three searches outlined below and 0.995 for the fourth). Initially, we chose households from the samples in the CBG's own microdata sampling area (PUMA). The search was stopped if the cost reached a cut-off value of 15.0. We chose the cut-off value such that $FT^2$ would be less than the 0.05 percentile of a Chi-square ($X^2$) distribution, as $FT^2$ is by design approximately Chi-square distributed with degrees of freedom (k) equal to the number of target columns minus one. If 200,000 search steps elapsed without reaching the cut-off value, we repeated the search using a progressively wider selection of microdata samples, first with all samples from PUMAs in the same county as the CBG, then all PUMAs in the same core-based statistical area (CBSA), and finally all PUMAs with a similar urban/rural percentage (with empirically chosen cut-offs for similarity). We parallelized the overall search simply by allocating the CBGs among the available local processors.

The resulting set of PUMS households provided the age, sex, race/ethnicity, employment industry, income, and school grade of all individuals in households for the synthetic population. We then generated additional individuals in group quarters based on counts from the preprocessing step. We assumed that only five types of GQs exist: institutional for ages under 18; institutional, civilian non-institutional, and military for ages 18 to 64; and institutional for ages 65 and over. We further assumed that a maximum of one GQ of each type is present in a CBG, and created one if at least 20 residents of the appropriate age and institutional status were present. GQ residents were assumed not to attend school outside the GQ, and non-institutionalized individuals ages 18 to 64 were assigned employment based on the results from the preprocessing step.

We assigned students to schools based on school sizes and CBG-to-school distances (calculated above). For each person in the synthetic population enrolled in preschool through 12th grade, we identified the closest school that offered the required grade level and had not yet been filled to capacity. We assigned the person to that school with 90% probability, or to the next closest with 10% probability, to represent students choosing schools outside of their home school districts [14]. If the five closest schools were full, we overfilled the closest (90%) or second closest (10%).

To generate workplaces and assign workers, we started with counts of commuters by industry category residing in each CBG of the synthetic population. Then we multiplied those counts by the industry-destination commute matrices (generated in Preprocessing) to obtain the number of people in each industry category working in-person at each destination CBG (as well as the number working outside the synthesis area). Within industries, the commuters from each origin CBG were shuffled and the required number assigned to each destination.

For each school, we assigned teachers by finding the CBG whose centroid is closest to the school's location, and selecting teachers randomly from the education-industry workers commuting to that destination. If that destination did not have enough education workers to account for the reported number of teachers, we made up the difference by drawing teachers first from CBGs in the same census tract, and then from CBGs in the same county. We followed a similar procedure to assign staff to group quarters; for each GQ in a CBG, we assumed that the number of staff is equal to $0.1 \times$ the number of residents for an institutional GQ or $0.02 \times$ the number of residents for a non-institutional GQ, and assigned staff from the workers commuting to its CBG (or nearby CBGs). For GQs, we drew staff from the "public administration" industry category (NAICS code 92) based on the simplifying assumption that the majority would be prison staff.

To generate all other workplaces, for each industry category, in each work destination (CBG), we drew employer sizes from the lognormal distribution estimated from CBP data for that location, until enough jobs were generated for the workers of that industry commuting to that destination. We then randomly assigned workers having the corresponding industry classification and commute destination to those workplaces, creating "placeholder" people as needed to represent workers commuting from outside the synthetic population. The latter will be part of a workplace network but will have no household or other microdata-based characteristics explicitly generated. For residents of the synthetic population whose work destination is outside the synthesis area, we generated single-person placeholder workplaces (these people have households but will have no explicitly generated workplace network).

Finally, we generated the network of regular (repeated) contacts as follows. First, we fully connected individuals within households. Then, to connect individuals to co-workers, classmates, etc., we algorithmically generated a network within each workplace, school, and GQ. For workplaces and schools, we used a stochastic block model (SBM) where the blocks are income categories and grade levels, respectively. In this model, the mean number of edges, $K_{ij}$, connecting a vertex in block $i$ to vertices in block $j$ is

$$K_{ij} = \begin{cases} (1-\alpha)K\frac{N_j}{N}, & i \neq j, \\ \alpha K + (1-\alpha)K\frac{N_j}{N}, & i = j \end{cases} \quad (3)$$

where $\alpha$ is the assortativity, $K$ is the mean degree of the network, and $N_j / N$ is the number of vertices in block $j$ divided by the total number of vertices (i.e., the probability of an edge connecting to a vertex in block $j$ in a fully random network). Thus, the probability of connecting individuals in different blocks was multiplied by $(1 - \alpha)$ relative to what it would be if all contacts were random. For generating the example networks below, we arbitrarily set $K$ to 8 (for workplaces) or 12 (for schools), and $\alpha$ to 0.9. Workers' incomes were derived from the synthetic individuals' microdata and arbitrarily split into two categories at the $40,000/year boundary (following the categories in the LEHD dataset). Students' grade levels were derived from synthetic individuals' microdata, while teachers were assigned to grade levels randomly in proportion to the number of students in each grade level at a given school. For group quarters, we connected individuals within each facility using the Watts-Strogatz (small world) algorithm with mean degree $K = 12$ and rewiring probability $\beta = 0.25$.

*Agent-based simulation*
We used a simple agent-based model to simulate disease transmission of a respiratory pathogen across the synthetic network for the Maryland area. Each agent in the model is an individual from the synthetic population, with a residence location (CBG) assigned on creation and a fixed (on the timescale of the simulation) set of home, work, and school network contacts. Individuals may be in one of four disease states: susceptible; exposed (but not yet infectious); infectious; and recovered. Disease parameters generally similar to COVID-19 and influenza were used, but no effect of seasonality nor of changes in behavior over time were assumed. During the simulation, an infectious individual has a fixed probability of exposing each of their network contacts to the contagion at a randomly chosen time during a (uniformly random) 8-12 day period. Afterwards, the individual enters the recovered state and is not susceptible to reinfection on the time scale of the simulation. Each secondarily exposed individual, if susceptible, enters the exposed state for 5 days and then becomes infectious. Individuals who live outside of the synthetic population (who belong to a work contact network but are missing a household network) were checked for exposure at regular (10-day) intervals, using the assumption that their probability of becoming exposed at home is equal to the average probability experienced by a person living within the population during that time. The same procedure was followed for individuals who work outside of the synthetic population (who belong to a household contact network but are missing a workplace network). At the start of the simulation, 300 randomly chosen individuals were set to an infectious state; the rest began in a susceptible state. A large initial seeding was used to reduce stochasticity at the beginning of the simulation and make it easier to understand the behavior of the model.

We conducted a set of simulations to examine the effect of network topology. For this purpose, a person's contact network consisted solely of home, work, and school contacts, and an infectious person had a 15% probability of infecting each contact. We measured cumulative infections over time in 10 replicate runs of 600 days on the synthetic network described above, and on identically sized networks created using common graph generation algorithms. These included the Barabasi-Albert ("scale-free") and Watts-Strogatz ("small world") algorithms with an integer-valued mean degree not exceeding the mean degree of the synthetic network, and the Erdos-Renyi ("random network") algorithm with mean degree equal to that of the synthetic network. Source code for the simulation is available at https://github.com/CDDEP-DC/DiCED-GlobS-ABM.

**Results**

To verify the synthetic population, we compared the mismatch obtained for each CBG to the critical value (at $\alpha = 0.05$) of the Freeman-Tukey statistic and to that of a random selection of microdata households. The synthetic population matched the census totals to a high degree of accuracy in nearly every CBG with respect to the counts (census columns) chosen as optimization targets. Fig 1A shows the result for the Maryland area; only 12 out of 5,825 CBGs (0.2%) were a poor match to the census data. We then calculated the mismatch between the synthetic population and census totals for an arbitrary selection of 31 census columns not included as targets in the optimization step. These counts showed a relatively poor match between the synthetic and real populations but were generally better than a random selection of households (Fig 1B). This result is consistent with other CO studies showing that correlations between different census columns are typically not strong enough to guarantee a good match for counts not specifically targeted for optimization [15]. Results for the Los Angeles area were similar (not shown).

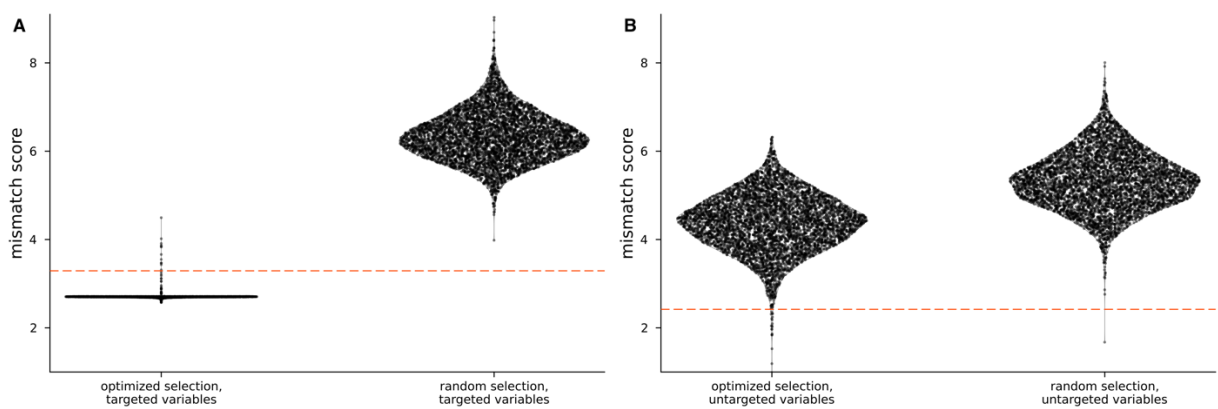

**Fig 1. Fit between census data and households selected by combinatorial optimization for each census block group (CBG) in the Maryland-area synthetic population.** (A) Comparison between an optimized selection of microdata households (left) and a random selection (right) showing the fit between the selected households and the census data with regard to census columns targeted during optimization. The mismatch score (y-axis) is $\ln(¼\, FT^2)$, where $FT^2$ is the Freeman-Tukey goodness-of-fit statistic (a lower score corresponds to a better fit). Each dot represents a single CBG. The red dotted line is the 0.95 percentile of the test statistic; a point above this line indicates that the selected households are a poor match to the census data for a CBG. Optimization was performed using simulated annealing as detailed in the main text. (B) The same comparison for an arbitrary set of census columns not targeted during optimization.

For verifying the synthetic contact network, no ground truth data were available. Instead, Table 4 shows network statistics for the Maryland- and Los Angeles-area networks compared to networks with similar mean degree generated using the Watts-Strogatz (small world), Barabasi-Albert (scale-free), "static" scale free[16], and Erdos-Renyi (random network) algorithms. The algorithmically generated networks were created with the same number of vertices as the Maryland-area network. Clustering and degree assortativity were calculated using Julia Graphs.jl version 1.8.0. Tendency to make hub (TMH) is a measure of vertex degree distribution from [17]. The vertex degree information index, $I_r(vd)$, is a measure of topological complexity from [18]. The synthetic networks showed higher clustering than the scale-free or random networks; higher degree assortativity than any of the algorithmically generated networks. TMH and $I_r(vd)$ were higher than the small world or random networks but lower than that of the scale-free network.

**Table 4. Synthetic contact networks (MD and LA) compared to algorithmically generated Watts-Strogatz (small world), Barabasi-Albert (scale-free), static scale-free[16], and Erdos-Renyi (random) networks.**

|  | MD network | LA network | Small world | BA scale-free | Static scale-free | Random |
|---|---|---|---|---|---|---|
| $\overline{D}$ | 8.48 | 8.53 | 8.0 | 8.0 | 8.48 | 8.48 |
| $\overline{C_I}$ | 0.307 | 0.375 | 0.276 | 2.14e-5 | 1.46e-5 | 9.83e-7 |
| $C$ | 0.156 | 0.19 | 0.263 | 1.16e-5 | 1.45e-5 | 9.29e-7 |
| $r$ | 0.385 | 0.4 | -0.0269 | -0.00322 | -5.22e-4 | -1.97e-4 |
| TMH | 12.3 | 12.6 | 8.22 | 41.0 | 36.4 | 9.48 |
| $I_r(vd)$ | 0.131 | 0.128 | 0.115 | 0.135 | 0.138 | 0.121 |

$\overline{D}$: mean degree; $\overline{C_I}$: mean local clustering coefficient; $C$: global clustering coefficient; $r$: degree assortativity coefficient; TMH: tendency to make hub[17]; $I_r(vd)$: vertex degree information index[18]

Based on simulation results, the Maryland-area synthetic network exhibited slower disease transmission than the Barabasi-Albert scale free network (even though the latter was generated with a slightly smaller number of mean connections). By contrast, the synthetic network exhibited faster transmission than a network of random connections (generated with the Erdos-Renyi algorithm, having the same number of mean connections). The Watts-Strogatz small world network produced substantially slower transmission than the random network (not shown).

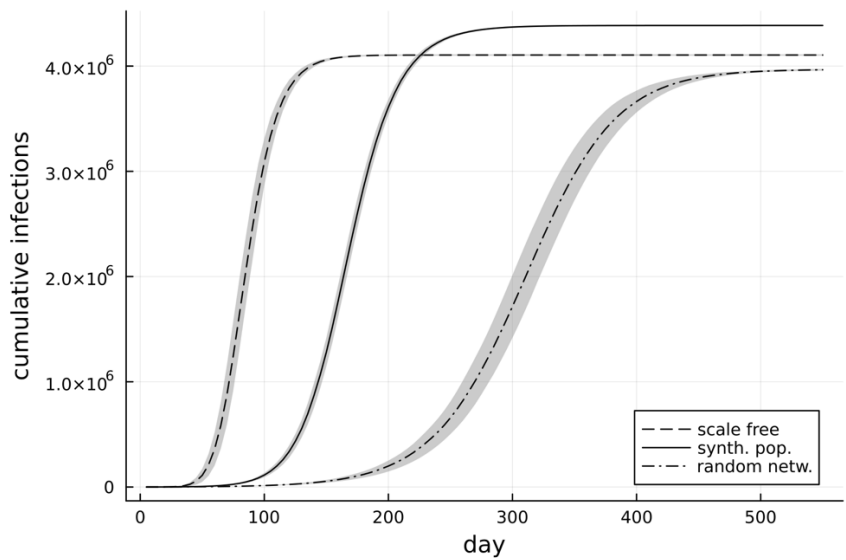

**Fig 2. Simulation of disease transmission showing the effect of different network topologies on transmission rate.** Cumulative infections over time in an individual-based simulation of disease transmission on the Maryland-area synthetic network, which has a mean of 8.48 connections per person (solid line, center), a Barabasi-Albert scale free network of identical size with a mean of 8 connections per person (dashed line, left), and an Erdos-Renyi random network of identical size with a mean of 8.48 connections (dot-dashed line, right). The shaded area represents 95% confidence intervals estimated from 10 simulation replicates.

**Discussion**

We presented a method that generates a synthetic population matching chosen properties of a real population at a fine (CBG-level) geographic scale. Our synthetic population was a good fit to the census data, even though we used only simple simulated annealing for the search and have not implemented the optimizations (e.g., fitting difficult targets first) described in [15]. The resulting contact network is realistic in that it approximately reflects the household composition and primary out-of-home activity of individuals in the region, as well as incorporating demographic details such as income and school grade that affect contact patterns.

The synthetic contact network has distinct properties that differentiate it from networks generated by simpler algorithms. It shows a high amount of clustering, moderately high topological complexity, and high vertex degree assortativity (i.e., individuals with many contacts tend to be connected to individuals with many contacts). This combination of features is not shared by any of the networks generated using the common graph-theoretic algorithms we tested. Furthermore, these differences appear to be relevant to epidemiological models of disease transmission; the synthetic contact network exhibits distinct transmission dynamics from that of the simpler networks in a simulated epidemic.

The software we provide can be used and extended by researchers to simulate any region of the US at the resolution available in the census data. It could potentially be applied to regions outside the US if the data can first be converted to a compatible format. Any feature of the population can be recreated, provided it is present in the census data and can be sampled by household from the microdata. For example, we were able to target household income and employment statistics to capture economic segregation both in the workplace and between neighborhoods.

The high level of detail our approach provides is potentially beneficial for many kinds of research questions. Our focus has been primarily in epidemiology, e.g., studying disparities in healthcare outcomes that might result from economically biased workplace closure policies, or predicting the effects of policy interventions. However, the feature-rich nature of the populations our method generates may also be useful for other areas of research, such as transportation[19], infrastructure and urban planning[20,21], public policy[22], consumer markets[23], and disaster response[24].

Many such questions are fruitfully studied by simulating interactions among individuals (i.e., an agent-based model). The demographic and geographic data contained in our synthetic populations provide an informed way to generate these interactions. In addition to the household, workplace, and school contact networks described above, sporadic or ephemeral contacts can be generated using the features of individuals in the population, e.g., between retail workers and people living or working in that location.

Furthermore, dynamic individual behavior is an important feature of many agent-based models, and previous research shows that changes in behavior are essential for understanding epidemic progression [25,26]. By providing details of interest about individuals in the simulation, our method allows behavior change to be modeled as interactions between external events and individual characteristics. This may be valuable for epidemic forecasting, for simulating the effects of behavioral interventions, and for understanding the effects of behavior-driven change in contact patterns (e.g., travel or attending public events).

*Data availability*
The data used in this study were extracted from resources available in the public domain. These are as follows: American Community Survey (ACS) 5-year summary and decennial census, data.census.gov; CBG TIGER/Line shapefiles, www.census.gov/geographies/mapping-files/time-series/geo/tiger-line-file.html; 5-year PUMS, www.census.gov/programs-surveys/acs/microdata.html; origin-destination (OD) employment statistics, lehd.ces.census.gov; County Business Patterns (CBP), www.census.gov/programs-surveys/cbp.html; US public schools, nces.ed.gov; urban/rural proportion by geographic area, mcdc.missouri.edu.

*Code availability*
The synthetic population software with source code is available at github.com/CDDEP-DC/GREASYPOP-CO. The agent-based simulation code is available at github.com/CDDEP-DC/DiCED-GlobS-ABM.

**Acknowledgments**

This work was funded by the Centers for Disease Control and Prevention (CDC) MInD-Healthcare Program (grant number 1U01CK000536), and by NSF PIPP Phase I: Evaluating the Effectiveness of Messaging and Modeling during Pandemics (grant number 2200256). The funders had no role in the design, analysis, decision to publish, or preparation of the manuscript.